\documentclass[twocolumn,pra,showpacs,superscriptaddress]{revtex4-1}
\usepackage{amssymb}
\usepackage{amsmath}
\usepackage{graphicx}
\usepackage{subfigure}
\usepackage{natbib}
\usepackage{epstopdf}
\usepackage{CJK}
\usepackage{amsfonts}
\usepackage{mathrsfs}
\usepackage{ulem}
\usepackage{color}
\usepackage[toc,page,title,titletoc,header]{appendix}

\begin{document}

\title{ Effective preparation and collisional decay of atomic condensate in
excited bands of an optical lattice}
\author{Yueyang Zhai}
\affiliation{School of Electronics Engineering and Computer Science, Peking University, Beijing 100871, China }
\author{Xuguang Yue}
\affiliation{School of Electronics Engineering and Computer Science, Peking University, Beijing 100871, China }
\author{Yanjiang Wu}
\affiliation{Department of Physics, Renmin University of China, Beijing 100190, China}
\author{Xuzong Chen}
\affiliation{School of Electronics Engineering and Computer Science, Peking University, Beijing 100871, China }
\author{Peng Zhang }
\email{pengzhang@ruc.edu.cn}
\affiliation{Department of Physics, Renmin University of China, Beijing 100190, China}
\author{Xiaoji Zhou}
\email{xjzhou@pku.edu.cn}
\affiliation{School of Electronics Engineering and Computer Science, Peking University, Beijing 100871, China }

\begin{abstract}
We present a method for the effective preparation of a Bose-Einstein condensate (BEC)
into the excited bands of an optical lattice via a standing-wave pulse sequence. With our method, the BEC can be prepared in either a single Bloch state in a excited-band, or a coherent superposition of states in different bands. Our scheme is experimentally demonstrated by preparing a $^{87}$Rb BEC into the $d$-band and the superposition of $s$- and $d$-band states of a one-dimensional optical lattice, within a few tens of microseconds. We further measure the decay of the BEC in the $d$-band state, and carry an analytical calculation for the collisional decay of atoms in the excited-band states. Our theoretical and experimental results consist well.
\end{abstract}

\maketitle

% Force line breaks with \\

%\author{Yueyang Zhai}

.

%*************************Introduction*******************************

\section{introduction}

Ultracold atomic gases in optical lattices have various applications in many
fields, including the quantum simulation of many-body systems and the
realization of quantum computation and high-precision atomic clock ~\cite{RMP06,RMP08,RMP11}. So far most of the experiments have been implemented in ground bands ($s$-bands) of optical lattices. Recently, ultracold
gases in the excited bands of optical lattices attract many attentions. It
is proposed that many interesting many-body phenomena, e.g., supersolid
quantum phases in cubic lattices~\cite{Scarola}, quantum stripe ordering in
triangular lattices~\cite{Wu}, orbital degeneracy~\cite{Lewenstein} can
appear in the ultracold atoms in the excited-band states. Nevertheless, the $%
d$- and $f$-band physics in optical lattices have remained experimentally
unexplored, except the bipartite square optical lattice~\cite{M}.

A common concern for the research of excited-band physics of ultracold gas
in an optical lattice is how to rapidly load the atoms into the high energy
bands without excitation or heating. So far several experimental techniques
have been developed for preparing ultracold atoms in the high energy bands.
These techniques include: (i) the coherent manipulation of vibrational bands
by stimulated Raman transitions~\cite{Muller}, (ii) using a moving lattice
to load a Bose-Einstein condensate (BEC) into a excited-band~\cite{Browaeys}%
, (iii) the population swapping technique for selectively exciting the atoms
into the $p$-band~\cite{Wirth1} or $f$-band~\cite{M} of a bipartite
square optical lattice. It is pointed out that, these approaches are
designed to transfer the atoms from the $s$-band to the excited bands.
Namely, to create an ultracold gas in the excited band of the optical lattice
with these approaches, one needs to first load the atoms into the $s$-band.
With the widely-used adiabatic loading approach, such a process takes
several tens of milliseconds.

In this paper, we develop a method for effective preparation of a
weakly-interacting Bose-Einstein condensate
(BEC) in the high energy bands of an
optical lattice. This scheme is based on our previous work for the rapid
loading of BEC into the ground state of an optical lattice via a
standing-wave laser pulse sequence~\cite{Liu,Xiong}. With our method, the
BEC can be \textit{directly} transferred from the ground state of the weak
harmonic trap into the excited band of the optical lattice with a
non-adiabatic process, which can normally be completed within several tens
of microseconds. Furthermore, in our scheme the BEC can be prepared in
either a single excited-band Bloch state or a coherent superposition of
Bloch states in different bands with the same quasi-momentum. As a
demonstration, we experimentally realize the effective preparation of a $^{87}$Rb
BEC into a $d$-band state, and the coherent superposition of $d$-band and $s$%
-band states of an one-dimensional (1D) optical lattice. The effectiveness
of our approach is further verified by the observations of the atomic Rabi
oscillations between states with different single-atom momentum. As shown
below, the fidelities of the preparation process in our experiments are as high
as 97\%-99\%.

As an application of our method, we experimentally investigate the decay
process of the $^{87}$Rb BEC, which is rapidly loaded in the $d$-band of the
1D optical lattice. It is well known that, when the ultracold atoms are
prepared in the excited-band state of an optical lattice, they can decay to
the states in the lower bands via inter-atomic collision. For the ultracold
gas prepared around the lowest-energy points of high energy bands, the
lifetime of the gas is mainly determined by the collisional decay. Such a
decay process was experimentally observed by N. Katz {\it et al.} in a
moving optical lattice~\cite{Katz} and theoretically studied with a
perturbative calculation by the same authors~\cite{Katz}. Nevertheless, to
the best of our knowledge, there is still lack of a first-principle
calculation for the collisional-decay rate. In this paper, based on the
scattering theory, we provide a first-principle calculation for the
collisional-decay process of ultracold gases in the excited bands, and
obtain the analytical expression of the decay rate. We compare our
theoretical result and the experimental observations, and find great
consistency.

The remainder of this manuscript is organized as follows. In Sec. II, we
introduce our method for effective preparation of BEC in the excited-band states.
Our experiment for the preparation of the BEC of $^{87}$Rb atoms is shown in
Sec. III. In Sec. IV we analytically calculate the rate of the collisional
decay of atoms in the $d$-band state, and compare our result with the
experimental observations. The main results are summarized and discussed in
Sec. V, while some details of our calculations are given in the appendix.

\begin{figure}[tbp]
\centering
\includegraphics[width=8.5cm]{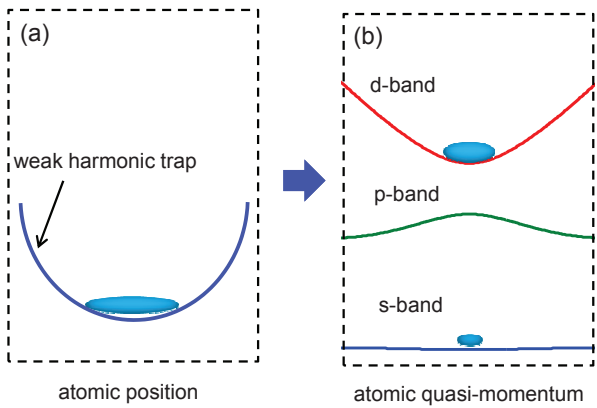}
\caption{(Color online) The system before and after the preparation process. (a) Before the
process, the BEC is confined in a weak harmonic trap. (b)
After the preparation process,
an optical lattice is turned on and the atoms in the condensate are
transferred into a single quasimomentum state, or the superposition of
quasimomentum states in different energy bands.}
\label{system}
\end{figure}

\section{Approach for effective preparation of BEC in excited bands}

Now we introduce our approach for
 rapidly preparing a BEC in the excited bands of an optical lattice. For simplicity, in this
section we only consider the case with 1D optical lattice. Our method
introduced here can be straightforwardly generalized to the systems with
two- or three-dimensional lattice.

In the ultracold gas of single-component bosonic atoms, a 1D optical lattice
can be created by two counter-propagating laser beams. In the presence of
the optical lattice, the single-atom Hamiltonian in the $x$-direction is
given by $(\hbar =1)$%
\begin{equation}
H_{x}=\frac{p_{x}^{2}}{2m}+V_{0}\cos ^{2}\left( \frac{x}{a}\pi \right) ,
\label{a}
\end{equation}%
with $m$ and $p_{x}$ the single-atom mass and momentum in the $x$-direction,
respectively. Here $V_{0}$ is the depth of the optical lattice and $a$ is
the lattice constant. According to the Bloch's theorem, the eigen-state of $%
H_{x}$ can be expressed as $|n,q\rangle \equiv u_{n,q}\left( x\right)
e^{iqx}/\sqrt{2\pi }$, with $n=s,p,d...$ the index of the energy band, and $%
q\in \lbrack -\pi /a,\pi /a)$ the quasi-momentum. Here the periodic Bloch
function $u_{n,q}\left( x\right) $ satisfies $u_{n,q}\left( x\right)
=u_{n,q}\left( x+a\right) .$ Using this property, it can be easily proved
that Bloch state $|n,q\rangle $ is the superposition of the plane waves $%
e^{ikx}/\sqrt{2\pi }$, with $k=q+2j\pi /a$ $(j=0,\pm 1,\pm 2,....)$. Namely,
$|n,q\rangle $ can be re-expressed as%
\begin{equation}
|n,q\rangle =\sum_{j=-\infty }^{+\infty }C_{nj}(q)|p_{x}=q+2j\pi /a\rangle ,
\label{c}
\end{equation}%
where $|p_{x}=k\rangle \equiv e^{ikx}/\sqrt{2\pi }$ is the eigen-state of $%
p_{x}$ with eigen-value $k$, and $C_{nj}$ is the superposition coefficient.

As shown in Fig.~\ref{system}, we suppose that before the preparation process,
there is no optical lattice in our system, and the atoms are condensed in
the single-atom ground state of the weak harmonic trap. We further
approximate such a state to be $|p_{x}=0\rangle $ with zero momentum. Our purpose is
to prepare the condensed atom in a given superposition state
\begin{equation}
|\Psi _{a}\rangle =\sum_{n}f_{n}|n,q_{0}\rangle .
\end{equation}%
According to Eq.~(\ref{c}), $|\Psi _{a}\rangle $ can be expressed as the
superposition of the states $|p_{x}=q_0+2j\pi /a\rangle $, i.e., we have
\begin{equation}
|\Psi _{a}\rangle =\sum_{j=-\infty }^{+\infty }d_{j}|p_{x}=q_{0}+2j\pi
/a\rangle ,  \label{d}
\end{equation}%
with $d_{j}=\sum_{n}f_{n}C_{nj}(q_{0})$.

We first consider a simple case where $|\Psi _{a}\rangle $ is the
superposition of the zero-quasi-momentum Bloch states in the
``even bands", i.e., the case with $q_{0}=0$ and $%
f_{p,f,h,...}=0$. In that case, with our method the preparation process is accomplished via alternating cycles of switching on (duty cycle) and off (off-duty cycle) the optical lattice. In these duty cycles, the atom experiences spatial potential $V_0\cos ^{2}\left( x\pi /a\right) $. Such a potential can induce the transition between the states $|p_{x}=2j\pi/a\rangle $ with different values of $j$. In the off-duty cycle, the atom is governed by the free-Hamiltonian $p_{x}^{2}/(2m)$. Thus, although there is no transition between different eigen-states of $p_{x}$, these states can gain different phase factors. Therefore, when the duty and off-duty cycles
are alternately applied to the atoms at state $|p_{x}=0\rangle $, the atoms
can be prepared to a superposition state of $|p_{x}=2j\pi /a\rangle $, i.e.,
a state with the form in Eq.~(\ref{d}). It is pointed out that, since the
initial atomic momentum is zero and the quasi-momentum is conserved in both
of the two cycles, in the preparation process the atomic state can only be the
superposition of the zero-quasi-momentum states in different energy bands.
Finally, when all the duty and off-duty cycles are completed, we \textit{instantaneously}
switch on
the optical lattice, and then the atoms are loaded in the optical lattice.

The above preparation approach can be mathematically described as follows. We
assume the preparation process includes $N_{C}$ duty cycles and $N_{C}$ off-duty
cycles, and the duration of the $l$th duty and off-duty cycle is $\tau _{l}$
and $\tau _{l}^{\prime }$, respectively. Thus, after the preparation process, the atomic state would be
\begin{equation}
|\Psi _{L}\rangle \equiv \prod_{l}e^{-i\frac{p_{x}^{2}}{2m}\tau _{l}^{\prime
}}e^{-i[\frac{p_{x}^{2}}{2m}+V_{0}\cos ^{2}\left( x\frac{\pi }{a}\right)
]\tau _{l}}|p_{x}=0\rangle .
\end{equation}%
Therefore, for a given target state $|\Psi _{a}\rangle $, the parameters $%
N_{C}$ and $\{\tau _{l},\tau _{l}^{\prime }\}$ can be determined via
maximizing the fidelity
\begin{equation}
F=\left\vert \langle \Psi _{L}|\Psi _{a}\rangle \right\vert ^{2}.  \label{e}
\end{equation}%
It is apparent that the value $1-F$ just describes the difference between
the realistic atomic state $|\Psi _{L}\rangle $ after the preparation
and the target state $|\Psi _{a}\rangle $, i.e., the error in the preparation process. When $F=1$ the atoms would be fully prepared in the state $|\Psi
_{a}\rangle $. It is pointed out that, for simplicity,
here we assume the optical lattice has the same intensity $V_0$ in all the duty cycles. In the practical cases, if it is necessary, the optical-lattice intensity can also be treated as a control parameter, and take different values
in different duty cycles. On the other hand, due to the selection rule, in the above process the
potential of the optical lattices in the duty cycles can only couple the initial state $%
|p_{x}=0\rangle $ with the states $|n,0\rangle $ with $n=s,d,g,...$. Thus,
the atoms can only be prepared into the states in these bands.

When the target state $|\Psi _{a}\rangle $ is the superposition of the
zero-quasi-momentum state in both ``even bands" and
``odd bands" (i.e., $q_{0}=0$ and $f_{p,f,h,\cdots}\neq 0$), the
preparation process can also be accomplished via a sequence of laser pulses.
Nevertheless, here one should use the laser pulses of optical lattices moving with a
velocity $v<\pi /(ma)$. Namely, the potential created in the $l$th duty
cycle should be proportional to $V_{l}\cos [(x-vt)\pi /a]$. The mechanism of
the preparation approach can be easily understood in the reference moving with
 velocity $v$. In that reference, the initial atomic state and
the target state in Eq.~(\ref{d}) become $|p_{x}=-mv\rangle $ and $%
\sum_{j=-\infty }^{+\infty }d_{j}|p_{x}=-mv+2j\pi /a\rangle $, respectively.
Thus, the pules in the duty cycles can induce the transition
between the states $|p_{x}=-mv+2j\pi /a\rangle $ with different values of $j$%
, and in the off-duty cycles these states can gain different phase factors.
Therefore, with the help of the sequence of the laser pules one can
prepare the atoms in the target state. It is easy to prove that, these pulses can induce the transition between the states in
any two bands, and thus the atoms can be prepared in the target state with
arbitrary coefficient $d_{j}$.

\begin{figure}[b]
\begin{center}
\includegraphics[width=8cm]{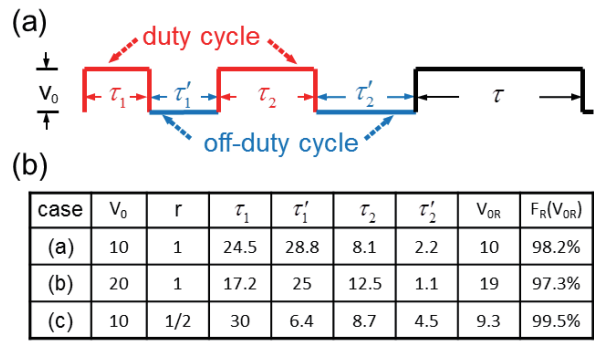}
\end{center}
\caption{(Color online) (a) Sequence of laser pulses in our experiments. (b) Table of the designed lattice depth $V_{0}$, the parameter $r$
for the target state $|\Psi _{a}\rangle $ in Eq.~(\protect\ref{psiaa}), the
durations $\protect\tau _{1,2}$ and $\protect\tau _{1,2}^{\prime }$ given by
the numerical maximizing of the fidelity $F$, the fitted value $V_{0R}$ of
the lattice depth and the fidelities $F_R(V_{0R})$ defined in Eq.~(\protect
\ref{fr}) for the preparation processes of the cases (a, b, c) in our
experiments.
Here $V_{0R}$ is given by the fitting of the
theoretical values of $N_0(\tau)/N$ given by Eq.~(\ref{g}) to the experimental measurements.
The units of $V_{0}$ and $V_{0R}$ are $E_{R}$, and the
units of $\protect\tau _{1,2}$ and $\protect\tau^{\prime}_{1,2}$ are $%
\mathrm{{\protect\mu}{s}}$.}
\label{process}
\end{figure}

Finally we consider the case with $q_{0}\neq 0$, i.e., the target state $%
|\Psi _{a}\rangle $ is the superposition of the Bloch states with non-zero
quasi-momentum. With our approach, we cannot prepare the atoms into such a
state in the lab reference. Nevertheless, as shown above, with
the laser pulses of optical lattices moving with velocity $-q_{0}/m$, the atoms
can be loaded into the state $|\Psi _{a}\rangle $ in the reference moving
with these pulses.

\section{Experimental results}

In our experiment, we first prepare a cigar shaped BEC of about $N=1\times
10^{5}$ $^{87}$Rb atoms in the $|F=2,m_{F}=2\rangle $ hyperfine ground state
in the Quadripole-Ioffe configuration trap, of which the axial frequency is $%
20\mathrm{Hz}$ and the radial frequency $220\mathrm{Hz}$~\cite{Liu,Zhou10}. The
1D optical lattice along the BEC's long axis ($x$-direction) can be created
by laser beams with wavelength $\lambda =2a=852\mathrm{nm}$, which is far
beyond the $^{87}$Rb transition line between $|F=2\rangle $ to $|F'=3\rangle $ .

In our experiments we prepare the atoms in the excited-band states of a 1D
optical lattice with depth $V_{0}$. We choose the target state to be%
\begin{equation}
|\Psi _{a}(V_{0})\rangle =\sqrt{1-r}|s,0;V_{0}\rangle +\sqrt{r}%
|d,0;V_{0}\rangle ,  \label{psiaa}
\end{equation}%
where $r$ is a real number and $|n,q;V_{0}\rangle $ is the Bloch state in
the $n$ band with quasi-momentum $q$. We perform the preparation processes for
the cases (a) $V_{0}=10E_{R},r=1$, (b) $V_{0}=20E_{R},r=1$ and (c) $%
V_{0}=10E_{R},r=1/2$, with $E_{R}=4\pi ^{2}\hslash ^{2}/(m\lambda ^{2})$. As
shown in above section, the preparation of the atoms into the state $|\Psi
_{a}(V_{0})\rangle $ can be accomplished via switching on and off the
standing-wave laser beam for the optical lattice. In our
experiment we choose $N_{C}=2$. Namely, the
preparation process is accomplished via two duty cycles and two off-duty cycles, as shown in
Fig.~\ref{process}(a). The laser pulses in the duty cycles are generated via a fast response
radio frequency switch together with a normal frequency source. As shown in
Sec. II, we determine the durations $\tau _{1,2}$ and $\tau _{1,2}^{\prime }$
for the duty and off-duty cycles by numerically maximizing the fidelity $F$
defined in Eq.~(\ref{e}). In Fig.~\ref{process}(b) we show the values of $\tau
_{1,2}$ and $\tau _{1,2}^{\prime }$ and the maximized fidelities given by
our numerical calculations. With the same calculation we also obtain the
finial state
\begin{eqnarray}
|\Psi _{L}(V_{0})\rangle &=&\left( e^{-i\frac{p_{x}^{2}}{2m}\tau
_{2}^{\prime }}e^{-i[\frac{p_{x}^{2}}{2m}+V_{0}\cos ^{2}\left( \frac{\pi x}{a%
}\right) ]\tau _{2}}\times \right.  \notag \\
&&\left. e^{-i\frac{p_{x}^{2}}{2m}\tau _{1}^{\prime }}e^{-i[\frac{p_{x}^{2}}{%
2m}+V_{0}\cos ^{2}\left( \frac{\pi x}{a}\right) ]\tau _{1}}\right)
|p_{x}=0\rangle  \notag \\
&\equiv &\sum_{n}f_{L,n}\left( V_{0}\right) |n,0\rangle  \label{psil}
\end{eqnarray}%
of the atoms after the preparation process in cases (a, b, c).

As shown in Sec. II, in the end of the preparation process we \textit{instantaneously} switch on the optical lattice, and hold it for time $\tau $. Then we switch off the laser beams and the magnetic trap, and image the expanding cloud after $30$\textrm{ms} time of flight using resonant probe light propagating along the $z$-axis. With this approach we can measure the number $N_{j}(\tau )$ of the atoms in the zero-momentum state  $|p_{x}=2j\pi/a\rangle $, while the population of the higher momentum states are negligible (less than $5\%$). As shown above, after the preparation process in cases (a, b, c), the atoms prepared in the state $|\Psi_{L}\left( V_{0}\right) \rangle $ in Eq.~(\ref{psil}). Thus, when the laser beams and the magnetic trap are switched off, the atomic state is $|\Psi _{\tau }\left( V_{0}\right) \rangle =\sum_{n}f_{L,n}\left( V_{0}\right) e^{-i\mathcal{E}_{n ,0}\tau }|n,0\rangle $. Here $\mathcal{E}_{n,0}$ is the eigen-energy of $H_{x}$ with respect to the state $|n,0\rangle $, respectively. Namely, we have $H_{x}|n,0\rangle =\mathcal{E}_{n,0}|n,0\rangle $. Therefore, the number $N_{j}(\tau)$ of atoms in the state $|p_{x}=2j\pi/a\rangle$ at time $\tau$ would be $N_{j}(\tau )=N|\langle p_{x}=2j\pi/a|\Psi _{\tau}\left( V_{0}\right) \rangle |^{2}$, and satisfies
\begin{eqnarray}
\frac{N_{j}(\tau )}{N}=P_j(\tau )\label{g}
\end{eqnarray}%
with the function $P_j(\tau )$ defined as
\begin{eqnarray}
P_j(\tau )\equiv \left\vert \sum_{n}f_{L,n}\left(
V_{0}\right) C_{n,j}e^{-i\mathcal{E}_{n ,0}\tau }\right\vert ^{2},  \label{ga}
\end{eqnarray}%
with $C_{n,j}$ defined in the above subsection. Eqs.~(\ref{g}) and (\ref{ga}) show that the atom number $N_{j}(\tau )$ oscillates with $\tau $.
\begin{figure}[tbp]
\begin{center}
\includegraphics[bb=47bp 219bp 520bp 620bp,clip,width=9cm]{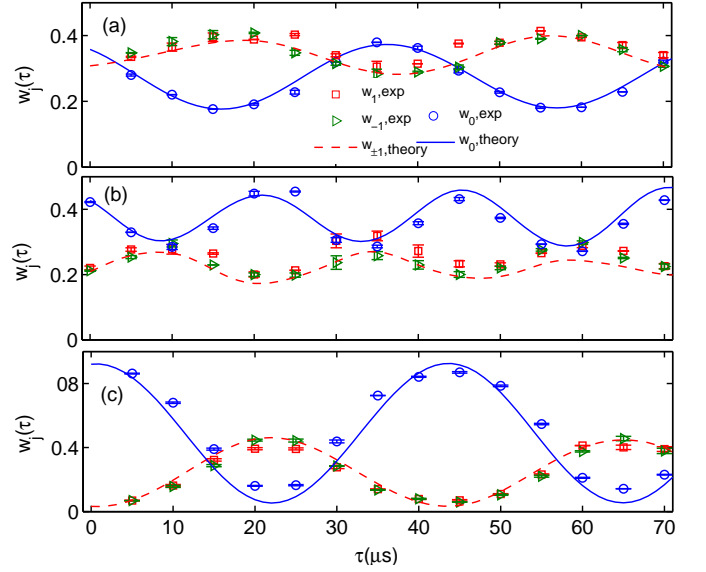}
\end{center}
\caption{(Color online) The relative population $N_j(\tau)/N$
($j=0,\pm 1$, it is denoted as $\mathrm{w}_j(\tau )$ in the figure)
for the states with $p_x=2j\pi/a$. Here $N_j(\tau)/N$ is plotted
as a function of the holding time $\tau $ for the cases
(a) $V_{0}=10E_{R},r=1$, (b) $V_{0}=20E_{R},r=1$ and (c) $V_{0}=10E_{R},r=1/2$.
We show the values of $N_0(\tau )/N$ (blue circles with error bar), $N_{1}(\tau )/N$
(red squares with error bar) and $N_{-1}(\tau )/N$ (green triangles with error bar) measured in our experiments and the ones
given by our numerical calculation with Eq. (\ref{g}) and the state $|\Psi _{\tau}(V_{0R})\rangle $
(blue solid line for $N_{0}(\tau)/N$ and red dashed line for $N_{\pm 1}(\tau)/N$),
with $V_{0R}$ given by the fitting of the
theoretical values of $N_0(\tau)/N$ given by Eq.~(\ref{g}) to the experimental measurements.}
%\caption{The zero-momentum-state population $P(\protect\tau )$ as a function
%of the holding time $\protect\tau $ for the cases (a) $V_{0}=10E_{R},r=1$,
%(b) $V_{0}=20E_{R},r=1$ and (c) $V_{0}=10E_{R},r=1/2$. Here we show the
%values of $P(\protect\tau )$ measured in our experiments (blue circle with
%error bar) and the ones given by the states $|\Psi _{L}(V_{0R})\rangle $
%from our numerical calculation (red solid line).}
\label{osc}
\end{figure}

In Fig.~\ref{osc} we illustrate the values of $N_j(\tau)/N$
($j=0,\pm 1$) of the state with $p_x=\pm2j\pi/a$
given by our experimental measurements. We also fit the theoretical values of $N_0(\tau)/N$ given by Eq.~(\ref{g}) to the experimental results. In our experiments, the values of $\tau _{1,2}$, $\tau _{1,2}^{\prime }$ and $a$ are controlled well. On the other hand, the relative accuracy of the control of $V_{0}$ is more than $90\%$. A relative error of $V_{0}$, which is in the order of one percent, may appear in our preparation process for each case. Due to this fact, in our calculations we use the experimental values of $\tau _{1,2}$, $\tau _{1,2}^{\prime }$ and $a$, and take $V_{0}$ as a fitting parameter. In Fig.~\ref{osc} we show the values of $N_j(\tau)/N$ given by
our theoretical calculation with Eq. (\ref{g}) and the lattice depth $V_{0R}$ given by the fitting calculation. It is shown that the theoretical curve fits well with
the experimental results. Therefore, in our experiments the atoms are successfully loaded in the state $|\Psi _{L}\left( V_{0R}\right) \rangle $. It is pointed out that,
the theoretical curves of $N_{1}(\tau)/N$ and $N_{-1}(\tau)/N$ are the same, while the experimental data differ by an amount of $5\%$. That difference may be caused by the imperfect alignment of the optical lattice along the long axis of the BEC.

In Fig.~2(b) we display the designed values $V_{0}$ and the realistic values $V_{0R}$ of the lattice depth in our experiments for the cases (a, b, c), and the realistic fidelities
\begin{equation}
F_{R}(V_{0R})=|\langle \Psi _{L}(V_{0R})|\Psi _{a}(V_{0R})\rangle |^{2}
\label{fr}
\end{equation}%
of the preparation processes in our experiments. It is shown that in cases (a)
and (b) where the target states are selected to be the single $d$-band Bloch state $|d,0;V_{0}\rangle $, the fidelities $F_{R}(V_{0R})$ of our
experimental preparation processes are as high as $98.2\%$ and $97.3\%$. In case (c) where the target state is the superposition state $(|s,0;V_{0}\rangle +|d,0;V_{0}\rangle )/\sqrt{2}$, the fidelity is $99.5\%$. According to these results, in all of our experiments with various target states and lattice depths, the preparation processes are successfully accomplished within several ten micro-seconds via our approach.

\section{Collisional decay and lifetime of BEC in excited
band}

In above sections, we show our approach to load the BEC to the excited band
of an optical lattice. As an application, we study decay process of the $%
^{87}$Rb BEC loaded in the $d$-band state with zero quasi-momentum of the 1D
optical lattice. It is well-known that, the atoms in the excited bands of an
optical lattice can decay to the lower bands via inter-atomic collision, and
the lifetime of these atoms is usually determined by this collisional decay.

In this section, we first give an analytical calculation for
the collisional-decay of the $^{87}$Rb BEC in our experiments. Then we
compare our theoretical result to our experimental measurements. The
quantitative agreement between them confirms our analytical result for the
collisional decay rate. Our result can be straightforwardly generalized to
other systems of weakly interacting BEC in the high energy bands of an
optical lattice.

We consider the ultracold bosonic atoms condensed in the $d$-band state with
zero quasi-momentum. When two atoms in the condensate decays to lower bands
via collision, they likely become thermal due to the large inter-band energy
gap. In the beginning of the collisional decay, these collisional products
are very rare. Thus, we can neglect the scattering between the thermal atoms
and the condensed ones, and only consider the collision of the atoms in the
condensate. Therefore, the decreasing of the density $n_{d}(t)$ of the
ultracold bosonic atoms condensed in the $d$ band can be described by the
master equation~\cite{ps}
\begin{equation}
\frac{dn_{d}\left( t\right) }{dt}=-Kn_{d}\left( t\right) ^{2}\,.  \label{ll}
\end{equation}%
Here the factor $K$ is given by
\begin{equation}
K=2\sum_{\left( n_{1},n_{2}\right) \neq (d,d)}\sigma \left(
n_{1},n_{2}\right) v,  \label{kk}
\end{equation}%
where $\sigma \left( n_{1},n_{2}\right) $ is the cross-section of the
two-atom inelastic collision, with the $i$th ($i=1,2$) atom in the $n_{i}$
band after the collision, and $v$ is the relative velocity of the two atoms
before collision. In Eq.~(\ref{kk}) the factor $2$ comes from the bosonic
statistics. Solving Eq.~(\ref{ll}), we obtain%
\begin{equation}
n_{d}\left( t\right) =\frac{n_{d}\left( 0\right) }{1+Kn_{d}\left( 0\right) t}%
,  \label{rr}
\end{equation}%
thus, the decay rate of our system can be defined as $\Gamma =Kn_{d}(0)$.

In the Appendix we calculate the cross-section $\sigma \left(
n_{1},n_{2}\right) $ for the system in our experiment, and obtain the result
($\hslash =1$)%
\begin{widetext}
\begin{equation}
\sigma \left( n_{1},n_{2}\right) v=\frac{4\pi a_{s}^{2}a^{2}}{m}\int dq\left[
\theta \left( 2\mathcal{E}_{d,0}-\mathcal{E}_{n_{1},q}-\mathcal{E}%
_{n_{2},-q}\right) \left\vert \int_{0}^{a}dxu_{n_{1},q}^{\ast }\left(
x\right) u_{n_{2},-q}^{\ast }\left( x\right) u_{d,0}^{2}\left( x\right)
\right\vert ^{2}\right] ,  \label{ss}
\end{equation}%
\end{widetext}with $m$ the mass of a single $^{87}$Rb atom and $a_{s}$ the
scattering length of two $^{87}$Rb atoms. Here $a$ is the lattice constant
of the optical lattice, the periodic Bloch function $u_{n,q}(x)$ and the
eigen-energy $\mathcal{E}_{n,0}$ of the Hamiltonian $H_{x}$ in the $x$%
-direction are defined in Sec. II. and Sec. III, respectively. In Eq.~(\ref%
{ss}) the $\theta $-function is defined as $\theta (x)=1$ for $x>0$ and $%
\theta (x)=0$ for $x<0$.

\begin{figure}[t]
\begin{center}
\includegraphics[bb=59bp 149bp 551bp 718bp,clip,width=9cm]{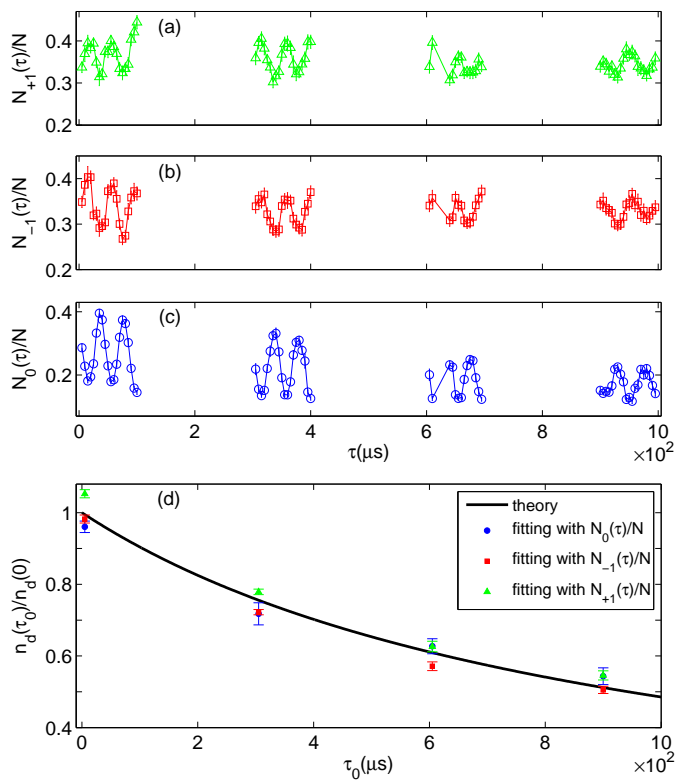}
\end{center}
\caption{(Color online) (a-c) The fraction $N_{j}(\tau )/N\ (j=0,\pm 1)$
given by our experimental measurements. Here $N_{j}(\tau )$ is the number
 of the atoms with $p_{x}\approx2j\pi/a $, and $N$ is the total atom number
in the condensate prepared in our experiment.
In our experiments we first perform the preparation  process for the case (a)
in Sec. III, and hold the optical lattice for time $%
\tau $. (d) The value of $n_{d}(\protect\tau _{0})/n_{d}(0)$ given
by the fitting of Eq.~(\ref{nn2}) with the experimental measurements
 of $N_{j}(\tau )/N\ (j=0,\pm 1,$ blue dots, green triangles and red suqares with error bars), and the one given by
 the theoretical calculation with Eq.~(\protect\ref{rr}) (black solid line).}
\label{decay}
\end{figure}

In our experiment, the collisional decay of the BEC in excited-band state is
observed via the following approach. We first perform the preparation process
for the case (a) in the above section, and prepare the $^{87}$Rb BEC at the
state $|\Psi _{L}(V_{0})\rangle =\sum_{n}f_{L,n}\left( V_{0}\right)
|n,0\rangle $. Here the depth $V_{0}$ of the optical lattice is $10E_{R}$
and the coefficients $f_{L,n}(V_{0})$ is given by the numerical calculation
in Eq.~(\ref{psil}), with $\tau _{1,2}$ and $\tau _{1,2}^{\prime }$ given in
Fig.~2(b). We have $|f_{d}|^{2}=98\%$. Then we hold the optical lattice for
time $\tau $ and measure the number $N_{j}(\tau )\ (j=0,\pm 1)$ of atoms in the
state $|p_{x}=2j\pi/a\rangle $. As shown in Fig.~4(a-c), we do the measurements for
the cases with $\tau =\tau _{0}+\tau ^{\prime }$ with $\tau _{0}=0,300%
\mathrm{\mu s},600\mathrm{\mu s},900\mathrm{\mu s}$ and $\tau ^{\prime }\in
(0,100\mathrm{\mu s})$.

Since $\tau ^{\prime }$ is much smaller than the characteristic time of the
collisional decay in our system, in time evolution of the BEC in the
interval $\tau \in (\tau _{0},\tau _{0}+\tau ^{\prime })$, the effect given
by the collisional decay can be neglected. Therefore, we have
the relation $N_{j}^d\left( \tau \right)/N=n_{d}\left( \tau
_{0}\right)/n_{d}\left(0\right)P_j(\tau)$, with $j=0, \pm 1$ and
the function $P_j(\tau)$ defined in Eq.~(\ref{g}). Here $N$
is the total number of the atoms in the condensate  prepared in our experiment.
Namely, when $\tau$ is large, $N$ is the summation of the atom number of the
remained condensate and the one of the product of the collisional decay.
In the above expression $N_{j}^d\left( \tau \right)$ is the
number of the atoms with $p_x=2j\pi/a$ in the remained condensate at time $\tau$.
Nevertheless, in our experiments, when $\tau$ is large the atoms in the remained condensate are mixed with some of the thermal atoms produced by the the collisional decay, which have the similar momentum with the condensed ones.
It is thus hard for us to exactly measure $N_{j}^d\left( \tau \right)$
for the cases with large $\tau$.
Therefore,
the number $N_{j}$ given by our measurement is actually the summation of the number of condensed atoms and the thermal atoms with $p_x\approx2j\pi/a$~\cite{nj}. Since the thermal atoms without quantum coherence do not attend the Rabi oscillation, we have
\begin{equation}
\frac{N_{j}\left( \tau \right) }{N}\approx \frac{n_{d}\left( \tau
_{0}\right) }{n_{d}\left( 0\right) }P_j(\tau)+n^t_j(\tau_0).  \label{nn2}
\end{equation}
with $j=0, \pm 1$ and $n^t_j(\tau_0)$ the density of the thermal atoms with $p_x\approx2j\pi/a$. It is pointed out that,
because $P_j(\tau)$ is an oscillating function of $\tau$,
the fraction $N_{j}\left( \tau \right)/N$ also oscillates with the time $\tau$. Physically speaking, that is because the quantum coherence is maintained in the remained
condensate. As shown in Fig.~4(a-c), such a behavior is clearly observed in our measurements.

We fit expression (\ref{nn2}) of $N_{j}(\tau )/N$ with the experimental measurements in each time interval $\tau _{0}<\tau <\tau _{0}+\tau ^{\prime }
$, and take $n_{d}(\tau _{0})/n_{d}(0)$ and $n^t_j(\tau_0)$ as the fitting parameter. In Fig.4(d)
we compare the value of $n_{d}\left( \tau _{0}\right) /n_{d}\left( 0\right) $
given by such a fitting, and the one given by Eq.~(\ref{rr}) with the factor
$K$ calculated from Eq.~(\ref{ss}) and $n_{d}\left( 0\right) =2.39\times
10^{14}~\mathrm{cm}^{-3}$. Here we approximate $n_{d}\left( 0\right) $ to be
the average atomic density of the condensate in our magnetic trap without
optical lattice. The good agreement between the theoretical and experimental
results confirms our analysis of the decay mechanisms and the calculations
of scattering amplitude. In particular, {\it all} of the oscillating amplitudes
of the curves $N_{0,\pm 1}(\tau)/N$ given by our measurements
quantitatively consist
with the condensate fraction $n_d(\tau_0)/n_d(0)$ given by our theoretical calculation.
This consistent shows that in our experiment
the quantum coherence are successfully maintained in the
un-decayed condensate, and does not exist in the decay products.

\section{conclusion}

In this paper we present a method for effective preparation of a BEC in excited bands of an optical lattice. With our approach the BEC can be
prepared in either a pure Bloch state in the excited band or the
superposition of Bloch states in different bands via the sequence of
standing-wave laser pulses. We experimentally demonstrate our method by
preparing the $^{87}$Rb BEC into the $d$-band state and the superposition of $%
s $- and $d$-band states of a 1D optical lattice within a few tens of
microseconds. We further measure the collisional decay process of the $d$%
-band BEC prepared in our experiment, and analytically derive the
collisional-decay rate atoms in the excited-band states. The experimental
and theoretical results consist well with each other. Our method and result
are helpful for the study of orbital optical lattice and simulation of
condensed matter physics.

\section{ACKNOWLEDGMENTS}

We thank Hui Zhai, Hongwei Xiong and Xinxing Liu for useful discussions. This work is supported
by National Natural Science Foundation of China under Grants No. 61027016,
61078026, 10934010, 11222430, and 11074305, and NKBRSF of China under Grants
No. 2011CB921501 and 2012CB922104.

\appendix%\appendixpage
\addcontentsline{toc}{section}{Appendices}\markboth{APPENDICES}{}
\begin{subappendices}

\section{the cross-section of inelastic collision between $d$-band
atoms}

In this appendix we calculate the cross-section $\sigma \left(
n_{1},n_{2}\right) $ of the inelastic collision between the two atoms in the
$d$-band state with zero momentum, and prove Eq.~(\ref{ss}). In the two-atom
scattering problem of our system, the total Hamiltonian is given by ($%
\hslash =1$)%
\begin{eqnarray}
H &=&-\frac{1}{m}\frac{\partial ^{2}}{\partial y^{2}}-\frac{1}{m}\frac{%
\partial ^{2}}{\partial z^{2}}+\sum_{i=1,2}\left[ -\frac{1}{2m}\frac{%
\partial ^{2}}{\partial x_{i}^{2}}+V(x_{i})\right]  \notag \\
&&+U\left( \vec{r}\right)  \notag \\
&\equiv &H_{0}+U\left( \vec{r}\right) ,  \label{h}
\end{eqnarray}%
with $\vec{r}=(x,y,z)$ the relative position of the two atoms, and $x_{i}$ $%
(i=1,2)$ the $x$-coordinate of the $i$th atom in the $x$-direction. Namely,
we have $x=x_{1}-x_{2}$. In Eq.~(\ref{h}), $V$ is the potential given by the
optical lattice in the $x$-direction, and $U\left( \vec{r}\right) $ is the
two-atom interaction potential. In this paper we model the inter-atomic
interaction with the Huang-Yang pseudo-potential
\begin{equation}
U\left( \vec{r}\right) =\frac{4\pi a_{s}}{m}\delta \left( \vec{r}\right)
\frac{\partial }{\partial r}(r\cdot ),  \label{u}
\end{equation}%
where $a_{s}$ the $s$-wave scattering length.

Here we calculate the cross-section with the approach in Sec. 3-e of Ref.~\cite{taylor}.
In Eq.~(\ref{h}), $H_{0}$ is defined as the free-Hamiltonian of the two
atoms without interaction. The eigen-state of $H_{0}$ can be written as%
\begin{eqnarray}
&&|\lambda ,n_{1},n_{2}\rangle \equiv  \notag \\
&&\frac{a}{\left( 2\pi \right) ^{2}}%
e^{ik_{y}y}e^{ik_{z}z}e^{iq_{1}x_{1}}e^{iq_{2}x_{2}}u_{n_{1},q_{1}}\left(
x_{1}\right) u_{n_{2},q_{2}}\left( x_{2}\right) ,\notag \\
\end{eqnarray}%
with $k_{y(z)}$ the two-atom relative momentum in the $y$-($z$-) direction
and $a$ the lattice constant of the optical lattice. Here $q_{i}$ and $n_{i}$
$(i=1,2)$ are the quasi-momentum and the quantum number for the energy band
of the $i$th atom, respectively. As shown in the maintext, $u_{n,q}(x)$ is the
periodic Bloch function of the $n$ band with quasi-momentum $q$. We further
define
\begin{equation}
\lambda =(k_{y},k_{z},q_{1},q_{2})
\end{equation}%
as the set of all the four quantum numbers. It is easy to prove that
\begin{eqnarray}
H_{0}|\lambda ,n_{1},n_{2}\rangle &=&\left( \frac{k_{y}^{2}+k_{z}^{2}}{m}+%
\mathcal{E}_{n_{1},q_{1}}+\mathcal{E}_{n_{2},q_{2}}\right) |\lambda
,n_{1},n_{2}\rangle  \notag \\
&\equiv &E_{\lambda ,n_{1},n_{2}}|\lambda ,n_{1},n_{2}\rangle ,
\end{eqnarray}%
where $\mathcal{E}_{n,q}$ is the single-atom energy associated to the $n$%
-band state with quasi-momentum $q$, and satisfies
\begin{equation}
\left[ -\frac{1}{2m}\frac{\partial ^{2}}{\partial x^{2}}+V(x)\right]
[e^{iqx}u_{n,q}(x)]=\mathcal{E}_{n,q}e^{iqx}u_{n,q}(x).  \label{eigen}
\end{equation}

Now we calculate the cross-section of the collision of two atoms in the $d$%
-band with zero quasi-momentum. According to the standard scattering theory,
the cross-section is defined with respect to a two-dimensional plane. Here
we assume the plane is spanned by the vectors $\hat{e}_{a}$ and $\hat{e}_{b}$.
They satisfy $\hat{e}_{a}\cdot \hat{e}_{b}=0$, and the $x$%
-component of $\hat{e}_{a(b)}$ is equal to $a$ (Fig.~5). To define the
cross-section, we should consider the incident wave packets
\begin{equation}
|\Psi _{\left( \kappa _{a},\kappa _{b}\right) }\rangle =\int
dk_{y}dk_{z}dq_{1}dq_{2}e^{-i\left( \kappa _{a}\hat{e}_{a}+\kappa _{b}\hat{e}%
_{b}\right) \cdot \vec{K}}\phi \left( \lambda \right) |\lambda ,d,d\rangle ,
\label{psiab}
\end{equation}%
where $\kappa _{a},\kappa _{b}$ are two integers, and $\lambda =\left(
k_{y},k_{z},q_{1},q_{2}\right) $ is the set of all the four quantum numbers.
Here $\phi \left( \lambda \right) \equiv \phi \left(
k_{y},k_{z},q_{1},q_{2}\right) $ is a normalized wave packet which sharply
peaks at the point $k_{y}=k_{y0},k_{z}=k_{z0},q_{1}=q_{10},q_{2}=q_{20}$,
and the vector $\vec{K}$ is defined as $\vec{K}\equiv \lbrack
(q_{2}-q_{1})/2,k_{y},k_{z}]$. With these assumptions, it is easy to prove
that the average two-atom relative position given by the wave function $%
|\Psi _{\left( \kappa _{a},\kappa _{b}\right) }\rangle $ is distributed in
the two-dimensional plane spanned by $\hat{e}_{a}$ and $\hat{e}_{b}$
(Fig.~5).

\begin{figure}[tbp]
\begin{center}
\includegraphics[width=6.5cm, ]{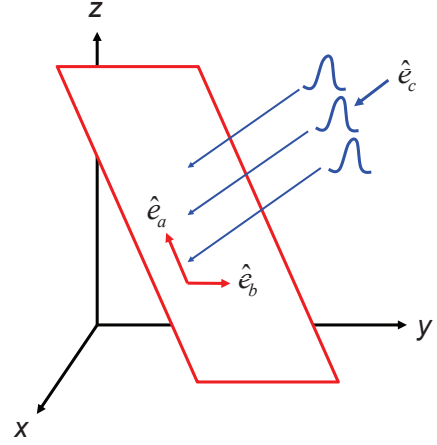}
\end{center}
\caption{The motion of the wave packets of the two-atom relative coordinate
through a two-dimensional plane.}
\label{decay}
\end{figure}

According to the scattering theory, the cross-section $\sigma \left(
n_{1},n_{2}\right) $ is defined as%
\begin{eqnarray}
&&\sigma \left( n_{1},n_{2}\right) =\left\vert \hat{e}_{a}||\hat{e}%
_{b}\right\vert \times  \notag \\
&&\sum_{\kappa _{a},\kappa _{b}}\int dk_{y}^{\prime }dk_{z}^{\prime
}dq_{1}^{\prime }dq_{2}^{\prime }\left\vert \langle \lambda ^{\prime
},n_{1},n_{2}|(S-1)|\Psi _{\left( \kappa _{a},\kappa _{b}\right) }\rangle
\right\vert ^{2},  \notag \\
&&  \label{aa}
\end{eqnarray}%
where $\lambda ^{\prime }=(k_{y}^{\prime },k_{z}^{\prime },q_{1}^{\prime
},q_{2}^{\prime })$ and $S$ is the S-operator with respect to the scattering
process, and satisfies%
\begin{eqnarray}
&&\langle \lambda ^{\prime },n_{1},n_{2}|S|\lambda ,d,d\rangle =\langle
\lambda ^{\prime },n_{1},n_{2}|\lambda ,d,d\rangle  \notag \\
&&-2\pi i\delta \left( E_{\lambda ,d,d}-E_{\lambda ^{\prime
},n_{1},n_{2}}\right) \langle \lambda ^{\prime },n_{1},n_{2}|T|\lambda
,d,d\rangle , \notag \\ \label{bb}
\end{eqnarray}%
with $T$ the associated T-operator.

Therefore, to obtain the scattering cross-section $\sigma \left(
n_{1},n_{2}\right) $, we should first calculate the T-matrix element $%
\langle \lambda ^{\prime },n_{1},n_{2}|T|\lambda ,d,d\rangle $. In our
experiments, since the scattering length of $^{87}$Rb atoms is much smaller
than the atomic de Broglie wavelength and the lattice constant $a$ of the
optical lattice, we can use the Born approximation
\begin{equation}
T\approx U\left( \vec{r}\right) .  \label{cc}
\end{equation}%
Using the Huang-Yang pseduo-potential in Eq.~(\ref{u}), we obtain%
\begin{equation}
\langle \lambda ^{\prime },n_{1},n_{2}|T|\lambda ,d,d\rangle =\frac{%
a^{2}a_{s}}{4m\pi ^{2}}A  \label{tr}
\end{equation}%
with the parameter $A$ defined as%
\begin{eqnarray}
A &\equiv &\int dx_{1}dx_{2}u_{n_{1},q_{1}^{\prime }}^{\ast }\left(
x_{1}\right) u_{n_{2},q_{2}^{\prime }}^{\ast }\left( x_{2}\right) \times
\notag \\
&&\delta \left( x_{1}-x_{2}\right) u_{d,q_{1}}\left( x_{1}\right)
u_{d,q_{2}}\left( x_{2}\right) .  \label{biga}
\end{eqnarray}%
To obtain the value of $A$, we first assume the length of the optical
lattice in the $x$-direction is $N_xa$. Then the direct calculation gives%
\begin{equation}
A=N_x\Gamma \left( q_{1},q_{2},q_{1}^{\prime },q_{2}^{\prime }\right) \delta
_{q_{1}+q_{2},q_{1}^{\prime }+q_{2}^{\prime }},
\end{equation}%
where the function $\Gamma \left( q_{1},q_{2},q_{1}^{\prime },q_{2}^{\prime
}\right) $ is defined as%
\begin{eqnarray}
&&\Gamma \left( q_{1},q_{2},q_{1}^{\prime },q_{2}^{\prime }\right)  \notag \\
&=&\int_{0}^{a}dxu_{n_{1},q_{1}^{\prime }}^{\ast }\left( x\right)
u_{n_{2},q_{2}^{\prime }}^{\ast }\left( x\right) u_{d,q_{1}}\left( x\right)
u_{d,q_{2}}\left( x\right) .\notag \\
\end{eqnarray}%
Here the Kronecker symbol is defined as $\delta _{ij}=1$ for $i=j$ and $%
\delta _{ij}=0$ for $i\neq j$. Therefore, for a slow-varying function $%
f(q_{1},q_{2})$, we have%
\begin{eqnarray}
&&\left( \frac{2\pi }{N_xa}\right) ^{2}\sum_{q_{1},q_{2}}Af(q_{1},q_{2})
\notag \\
&=&\frac{1}{N_x}\left( \frac{2\pi }{a}\right) ^{2}\sum_{q_{1}}\Gamma \left(
q_{1},q_{1}^{\prime }+q_{2}^{\prime }-q_{1},q_{1}^{\prime },q_{2}^{\prime
}\right) f(q_{1},q_{1}^{\prime }+q_{2}^{\prime }-q_{1})  \notag \\
&=&\frac{2\pi }{a}\int dq_{1}\Gamma \left( q_{1},q_{1}^{\prime
}+q_{2}^{\prime }-q_{1},q_{1}^{\prime },q_{2}^{\prime }\right)
f(q_{1},q_{1}^{\prime }+q_{2}^{\prime }-q_{1})  \notag \\
&=&\frac{2\pi }{a}\int dq_{1}dq_{2}\Gamma \left( q_{1},q_{2},q_{1}^{\prime
},q_{2}^{\prime }\right) f(q_{1},q_{2})\times  \notag \\
&&\delta \left[ \left( q_{1}^{\prime }+q_{2}^{\prime }\right) -\left(
q_{1}+q_{2}\right) \right] .  \label{biga2}
\end{eqnarray}%
Here we have used the relation%
\begin{equation}
\left( \frac{2\pi }{N_xa}\right) \sum_{q_{1}}=\int dq_{1}
\end{equation}%
which is applicable in the limit $N_x\rightarrow \infty $. The result in Eq.~(%
\ref{biga2}) implies%
\begin{equation}
A=\frac{2\pi }{a}\Gamma \left( q_{1},q_{2},q_{1}^{\prime },q_{2}^{\prime
}\right) \delta \left[ \left( q_{1}^{\prime }+q_{2}^{\prime }\right) -\left(
q_{1}+q_{2}\right) \right] .  \label{bigar}
\end{equation}%
Substituting Eq.~(\ref{bigar}) into Eq.~(\ref{tr}), we finally obtain the
element of T-matrix%
\begin{eqnarray}
&&\langle \lambda ^{\prime },n_{1},n_{2}|T|\lambda ,d,d\rangle  \notag \\
&=&\frac{aa_{s}}{2m\pi ^{2}}\Gamma \left( q_{1},q_{2},q_{1}^{\prime
},q_{2}^{\prime }\right) \delta \left[ \left( q_{1}^{\prime }+q_{2}^{\prime
}\right) -\left( q_{1}+q_{2}\right) \right] . \notag \\ \label{tmr}
\end{eqnarray}

Substituting Eq.~(\ref{tmr}) into Eqs. (\ref{bb}, \ref{psiab}) and Eq.~(\ref%
{aa}), we can obtain the scattering cross-section $\sigma \left(
n_{1},n_{2}\right) $. The straightforward calculation gives%
\begin{eqnarray}
&&\sigma \left( n_{1},n_{2}\right) =\frac{4\pi a_{s}^{2}a^{2}}{m}\int
dqdk_{y}dk_{z}dq_{1}dq_{2}\frac{\left\vert \phi \left( \lambda \right)
\right\vert ^{2}}{\left\vert \partial E_{\lambda ,d,d}/\partial k_{\parallel
}\right\vert } \notag \\
&&\times \theta \left( \mathcal{E}_{d,q_{1}}+\mathcal{E}_{d,q_{2}}-\mathcal{E}%
_{n_{1},\left( q_{1}+q_{2}\right) /2+q}-\mathcal{E}_{n_{2},\left(
q_{1}+q_{2}\right) /2-q}\right)  \notag \\
&&\times \left\vert \Gamma \left( q_{1},q_{2},\left( q_{1}+q_{2}\right) /2+q,\left(
q_{1}+q_{2}\right) /2-q\right) \right\vert ^{2}, \notag\\ \label{sigmar}
\end{eqnarray}%
where $k_{\parallel }$ is defined as $k_{\parallel
}=(q_{2}-q_{1},k_{y},k_{z})\cdot \hat{e}_{c}$. Here $\hat{e}_{c}$ is the
unit vector perpendicular to the plane spanned by $\hat{e}_{a}$ and $\hat{e}%
_{b}$, i.e., we have $|\hat{e}_{c}|=1$ and $\hat{e}_{c}\cdot \hat{e}_{a}=%
\hat{e}_{c}\cdot \hat{e}_{b}=0$. In Eq.~(\ref{sigmar}) the derivative $%
\partial E_{\lambda _{0},d,d}/\partial k_{\parallel }$ is taken for fixed
values of $q_{20}+q_{10}$ and $(q_{2}-q_{1},k_{y},k_{z})\cdot \hat{e}_{a}$
and $(q_{2}-q_{1},k_{y},k_{z})\cdot \hat{e}_{b}$. To obtain Eq.~(\ref{sigmar}%
), we have used the relation%
\begin{eqnarray}
&&\sum_{\kappa _{a},\kappa _{b}}e^{-i\left( \kappa _{a}\hat{e}_{a}+\kappa
_{b}\hat{e}_{b}\right) \cdot \vec{J}}=\frac{4\pi ^{2}}{\left\vert \hat{e}%
_{a}||\hat{e}_{b}\right\vert }\delta \left( \vec{J}\cdot \frac{\hat{e}_{a}}{%
\left\vert \hat{e}_{a}\right\vert }\right) \delta \left( \vec{J}\cdot \frac{%
\hat{e}_{b}}{\left\vert \hat{e}_{b}\right\vert }\right) ,  \notag \\
&&
\end{eqnarray}%
with $\vec{J}$ a vector in the three-dimensional space. Moreover, using the
fact that $|\phi \left( \lambda \right) |$ is sharply peaked at $\lambda
=\lambda _{0}$ and the relations%
\begin{eqnarray}
&&\int dk_{y}dk_{z}dq_{1}dq_{2}\left\vert \phi \left( \lambda \right)
\right\vert ^{2}=1, \\
&&q_{10}\approx 0,q_{20}\approx 0,k_{y0}\approx 0,k_{z0}\approx 0,
\end{eqnarray}%
we can further simplify Eq.~(\ref{sigmar}) and obtain the finial expression
for the cross-section of inelastic collisions between two $d$-band atoms
with zero quasi-momentum:
\begin{eqnarray}
&&\sigma \left( n_{1},n_{2}\right) =\frac{4\pi a_{s}^{2}a^{2}}{m\left\vert
\left. \partial E_{\lambda ,d,d}/\partial k_{\parallel }\right\vert
_{\lambda =\lambda _{0}}\right\vert }\times  \notag \\
&&\int dq\left[ \theta \left( 2\mathcal{E}_{d,0}-\mathcal{E}_{n_{1},q}-%
\mathcal{E}_{n_{2},-q}\right) \left\vert \Gamma \left( 0,0,q,-q\right)
\right\vert ^{2}\right] .  \notag \\
&&  \label{sigmarr}
\end{eqnarray}

Now we prove Eq.~(\ref{ss}). To this end, we need to calculate the component
$v$ of the two-atom relative velocity along the direction which is
perpendicular to the plane spanned by $\hat{e}_{a}$ and $\hat{e}_{b}$. It is
apparent that $v$ is defined as
\begin{equation}
v=\left\vert \langle \Psi _{\left( \kappa _{a},\kappa _{b}\right) }|-i\frac{%
\vec{\nabla}_{r}}{m}|\Psi _{\left( \kappa _{a},\kappa _{b}\right) }\rangle
\cdot \hat{e}_{c}\right\vert ,
\end{equation}%
where $\vec{\nabla}_{r}=(\partial /\partial x,\partial /\partial y,\partial
/\partial z)$, and the derivative is taken for fixed values of $x_{1}+x_{2}$%
. With the expression (\ref{psiab}) of $|\Psi _{\left( \kappa _{a},\kappa
_{b}\right) }\rangle $, it is easy to see that $v$ is independent on the
values of $\kappa _{a}$ and $\kappa _{b}$. To calculate the value of $v$, we
need the expressions of the periodic Bloch function $u_{d,q}\left( x\right) $
and the single-atom energy $\mathcal{E}_{n,q}$. We find that Eq.~(\ref{eigen}%
) for $u_{d,q}\left( x\right) $ and $\mathcal{E}_{n,q}$ can be simplified to%
\begin{equation}
\left[ -\frac{1}{2m}\frac{\partial ^{2}}{\partial x^{2}}+V(x)+h_{1}\right]
u_{n,q}(x)=\mathcal{E}_{n,q}u_{n,q}(x)
\end{equation}%
with $h_{1}=-i(q/m)\left( \partial /\partial x\right) $. Since the $|\phi
\left( \lambda \right) |$ is sharply peaked at $%
(q_{1},q_{2})=(q_{10},q_{20})\approx (0,0)$, we can treat the term $h_{1}$
in the above equation as a perturbation. The second-order perturbation
calculation gives the result%
\begin{equation}
v=\left\vert \left. \partial E_{\lambda ,d,d}/\partial k_{\parallel
}\right\vert _{\lambda =\lambda _{0}}\right\vert .  \label{vr}
\end{equation}%
Using Eq.~(\ref{sigmarr}) and Eq.~(\ref{vr}), we immediately obtain the
result in Eq.~(\ref{ss}).

\end{subappendices}


\begin{references}

\bibitem{RMP06}O. Morsch, M. Oberthaler, Rev. Mod. Phys. {\bf 78}, 179 (2006).

\bibitem{RMP08} I. Bloch, J. Dalibard, and W. Zwerger, Rev. Mod. Phys. {\bf 83}, 331 (2011).

\bibitem{RMP11} A. Derevianko and H. Katori, Rev. Mod. Phys. {\bf 80}, 885 (2008).

\bibitem{Scarola} V.W. Scarola and S. Das Sarma, Phys. Rev. Lett. {\bf 95}, 033003 (2005).

\bibitem{Wu}  C. Wu et al., Phys. Rev. Lett. {\bf 97}, 190406 (2006).

\bibitem{Lewenstein}  M. Lewenstein and W. V. liu, Nat. Phys. {\bf 7}, 101(2011).

\bibitem{M} M. \"{O}lschl\"{a}ger, G. Wirth, and A. Hemmerich, Phys. Rev. Lett. {\bf 106}, 015302 (2011).
\bibitem{Muller} T. M\"uller, S. F\"olling, A. Widera, and I. Bloch, Phys. Rev. Lett. {\bf 99}, 200405 (2007).

\bibitem{Browaeys} A. Browaeys, H. H\"affner, C. McKenzie, S. L. Rolston, K. Helmerson, and W. D. Phillips, Phys. Rev. A {\bf 72}, 053605(2005).
\bibitem{Wirth1} G. Wirth, Molschlager, and A. Hemmerich, Nature Phys. {\bf 7}, 147(2011).
%\bibitem{Wirth2} Molschlager, G. Wirth, and A. Hemmerich, Phys. Rev. Lett. {\bf 106}, 015302(2011).
\bibitem{Liu} X. Liu, X. J. Zhou, W. Xiong, T. Vogt, X. Z. Chen, Phys. Rev. A  {\bf 83}, 063402 (2011).
\bibitem{Xiong} W. Xiong, X. Yue, Z. Wang, X. J. Zhou, X. Z. Chen, Phys. Rev. A  {\bf 84}, 043616(2011).
\bibitem{Katz} N. Katz, E. Rowen, R. Ozeri, and N. Davidson, Phys. Rev. Lett. {\bf 95}, 220403 (2005).
\bibitem{Zhou10} X. J. Zhou, F. Yang, X. G., T. Vogt, and X. Z. Chen, Phys. Rev. A {\bf 81}, 013615 (2010).

\bibitem{ps} C. J. Pethick and H. Smith, \textit{Bose-Einstein Condensation in Dilute Gases}, Cambridge University Press, New York,
2002.

\bibitem{nj} For the cases with small $\tau$, e.g., the cases in Sec. III, we also have $N_{j}(\tau)\approx N_{j}^d\left( \tau \right)$ because in these cases the decay effect is negligible.

\bibitem{taylor} J. R. Taylor, \textit{Scattering Theory}, Wiley, New York,
1972.


\end{references}
\end{document}